\documentclass[twoside]{article}

\usepackage{amssymb}
\usepackage{amsmath}
\usepackage{amsthm}%% proof etc
\usepackage{algorithm}
\usepackage[noend]{algorithmic}
\usepackage{subcaption}

\usepackage{hyperref} %% inclusion after algorithm: mandatory
\usepackage{url}

\usepackage{boxedminipage}
\usepackage{graphicx}%%[draft] : do not embed figs/picts
\usepackage{float}
\usepackage{rotating}

\usepackage{xspace}
\usepackage{nicefrac}

\usepackage{comment}
\usepackage{soul}
\usepackage{verbatim}%%block comment

\usepackage{csquotes} % \textquote{}

\usepackage[T1]{fontenc}    % for accents
\usepackage[utf8]{inputenc} % for coding
\usepackage{xparse}   % \NewDocumentCommand
\usepackage{pifont}
\usepackage{marvosym}

\usepackage[table,usenames,dvipsnames]{xcolor}

\usepackage{fullpage}
\usepackage{arydshln}
\usepackage{comment}

\usepackage{lineno}
\newtheorem{remark}{Remark}

\input{macros-abs.sty}
\input{macros-local.sty}

\newif\ifAISTATS
\AISTATSfalse

\begin{document}
%\twocolumn[

\title{Efficient computation of the volume of a polytope in high-dimensions using Piecewise Deterministic Markov Processes}
\author{ Augustin Chevallier, Frédéric Cazals,  Paul Fearnhead }
%\aistatsaddress{ Lancaster University \And  Inria and Université Côte d'Azur \And Lancaster University } ]

\maketitle

\begin{abstract}

Computing the volume of a polytope in high dimensions is
computationally challenging but has wide applications. Current
state-of-the-art algorithms to compute such volumes rely on efficient
sampling of a Gaussian distribution restricted to the polytope, using
e.g. Hamiltonian Monte Carlo. We present a new sampling strategy that
uses a Piecewise Deterministic Markov Process. Like Hamiltonian Monte
Carlo, this new method involves simulating trajectories of a
non-reversible process and inherits similar good mixing
properties. However, importantly, the process can be simulated more
easily due to its piecewise linear trajectories — and this leads to a
reduction of the computational cost by a factor of the dimension of
the space. Our experiments indicate that our method is numerically
robust and is one order of magnitude faster (or better) than existing
methods using Hamiltonian Monte Carlo. On a single core processor, we
report computational time of a few minutes up to dimension 500.

\end{abstract}

%\noindent{\bf Keywords.} polytope volume, randomized algorithms,
%multi-phase Monte Carlo, Hamiltonian Monte Carlo, numerics,
%multi-precision arithmetic, exact predicates.

%\clearpage
%\input{aistats-main-v0.tex}

%Fast MCMC sampling algorithms on polytopes

\section{Introduction}
%%i%%%%%%%%%%%%%%%%%%%%%%%%%%%%%%%%%%%%%%%%%%%%%%%%%%%%%%%%%%%%%%%%%%%%%%%%%%%%%%%

\subsection{Volume of polytopes}

High dimensional integration and sampling is a pervasive challenge in
modern science. A subproblem of prime importance consists of computing
the volume of a polytope, a bounded convex set of $\mathbb{R}^d$
defined as the intersection of a fixed set of half spaces (H-polytope)
or the convex hull of a finite set of vertices (V-polytope). This challenge, of estimating the volume of a set, is closely related to problems in Statistics of computing marginal likelihoods \cite{fong2020marginal} or Bayes factors for comparing competing models \cite{gelman1998simulating,friel2012estimating}. It is also related to problems in Physics of estimating partition functions and free energies \cite{christ2010basic}. 

Other applications appear across a range of disciplines. For example,
in systems biology, the study of genome wide metabolic models based on
systems of ODEs requires studying polytopes defined as the intersection
between a hyperplane and the null space of a {\em stoichiometry}
matrix \cite{haraldsdottir2017chrr,chalkis2021geometric}.
In robotics, the computation of {\em reachable sets} for time-varying linear systems
is based upon special polytopes called zonotopes \cite{althoff2011reachable}.
In finance, the cross sectional score of a portfolio
is defined from the intersection between a simplex (representing assets)
and hyperplanes or ellipsoids \cite{cales2018practical}.
In artificial intelligence, ReLU networks can be characterized by
the conjunction of a set of linear inequalities which define a
polytope in the input domain known as the activation condition \cite{puasuareanu2020probabilistic}.
In a related vein, the regularization of neural networks
using piecewise affine functions involves the evaluation
of the volume of polytopes defined by intersections of these hyperplanes
\cite{robinson2021approximate}.
\medskip

Complexity-wise, computing the volume is \#-P hard irrespective of the
representation of the polytope (H-polytope or V-polytope)
~\cite{dyer1988complexity}. 
Intuitively, any deterministic algorithm using a polynomial number of
points and computing the corresponding convex hull omits an
exponentially large fraction of the volume. 
This observation prompted the development of approximation algorithms
delivering $(\eps,\delta)$ approximations, that is volume
approximations within a factor $1+\eps$ with a probability at least
$1-\delta$~\cite{barany1987computing,levy1997flavors}.
The complexity of such algorithms is measured by the number of calls
to an oracle stating whether a point is inside the polytope, or
alternatively computing the intersection between a line and the
polytope boundary. Remarkably, over the years, the
complexity has been lowered from $\ostar{d^{23}}$
\cite{dyer1991random} to $\ostar{d^{5}}$ \cite{kannan1997random} using
a sequence of balls intersecting the convex body, then from
$\ostar{d^{4}}$ \cite{lovasz2006simulated} to
$\ostar{d^{3}}$~\cite{cousins2015bypassing} using a sequence of smooth
probability densities restricted to the convex body. 
The reader is referred to \cite{lee2018convergence} for the full history.

Importantly, it should be stressed that the backbone of such
algorithms is a telescoping product (Section
\ref{sec:vol-estim-algos}) whose individual terms are the ratio of the
integrals (i) of Gaussians on the polytope, or (ii) of the identity
function integrated on the intersection between the polytope and a
convex body. The former strategy is theoretically faster but the
latter puts less burden on the samplers--see next section, since
uniform distributions are used.  \toblack

While the previous works are remarkable from the theoretical
standpoint, it is difficult to turn these algorithms into
effective implementations. This task is indeed challenging due to possibly huge constants in the complexities and un-realistic worst
cases.  This state of affairs recently motivated the development of
strategies relaxing the theoretical guarantees, based on novel
algorithmic and statistical techniques
\cite{cousins2016practical,chalkis2019practical,chevallier2022improved,chalkis2020volesti}.
While these methods are not provably correct in general, their
performances in terms of accuracy and running time have proven satisfactory.

\subsection{Samplers}
The most recent volume computation algorithms mentioned above rely on a procedure that samples a Gaussian distribution restricted to the polytope. 
The standard way to sample a probability distribution $\pi$ is to build a Markov
chain with invariant distribution~$\pi$. 

For sampling from a Gaussian distribution restricted to a compact region, a number of Markov chains have been proposed, including the Ball Walk sampler \cite{Lovasz-mixing-conductance}, the Hit and Run and coordinate Hit and Run samplers \cite{Lovasz99-HAR-mixes-fast,Lovasz03hit-and-runis,Lovasz04hit-and-run-corner,haraldsdottir2017chrr}, and Hamiltonian Monte Carlo \cite{pakman2014exact,chevallier2022improved}.

The Ball Walk is a simple Markov chain where the next point is proposed in a ball centered around the current point. If the proposed point is outside of the polytope, or if it rejected by Metropolis-Hasting, the chain stays at the current point.

By comparison, Hit and Run is a rejection free algorithm: a line
passing through the current point is chosen at random, and the next
point is sampled from the intersection of this line and the inside of
the polytope. While this Hit and Run mixes faster than the Ball Walk
sampler, it can get stuck in
corners~\cite{Lovasz04hit-and-run-corner,chevallier2022improved}. The
computational cost of Hit and Run can be reduced by restricting the
lines chosen to axis--coordinate Hit and Run.

Hamiltonian Monte Carlo can also be viewed as a rejection free method. It produces trajectories that are based on Hamiltonian dynamics for a physical system whose potential energy is defined by the Gaussian distribution within the polytope, and reflects the trajectory if it hits the boundary of the polytope. This sampler does not get stuck in corners like Hit and Run and has a very good mixing. However, unlike Hit and Run, the computation of the intersection of the trajectories and the boundary requires inverse trigonometric functions, and the computational cost cannot be reduced like Coordinate Hit and Run.

Another sampler is based on the billiard walk \cite{gryazina2014random}, which can be seen as a special case of HMC for uniform distributions, and is amenable to a computational complexity reduction similar to the one of coordinate Hit and Run \cite{chalkis2019practical}. This sampler can only target uniform distributions and not Gaussians.
\toblack

Finally, a new class of samplers based on Piecewise Detereministic Markov Processes (PDMPs) has emerged in the computational statistics community \cite{PDMPGeneral}. 
Recall that a PDMP is a deterministic process in-between random jump events, which occur at a certain rate.
Hamiltonian Monte Carlo can actually be seen as a PDMP. However, Hamiltonian Monte Carlo relies entirely on the Hamiltonian flow to preserve the target measure, while most of the PDMP samplers  developed to date have straight paths. The target measure is then preserved by changing direction at opportune random times. These methods are non reversible, which prevents the diffusive behavior of reversible chains, and have good properties in high dimension \cite{bierkens2019highdimensional}.

\subsection{Contributions}

In this work we introduce the use  of a PDMP, the Bouncy Particle Sampler \cite{BPSOriginal}, to sample from a Gaussian restricted to a polytope within algorithms for calculating the volume of the polytope. Like Hamiltonian Monte Carlo, the Bouncy Particle Sampler has good mixing properties. Moreover, we show how we can re-use many calculations so that it has a much lower computational overhead. The key idea is that the main computation in sampling from a polytope is checking whether and when a trajectory hits the boundary of the polytope. For the Bouncy Particle Sampler, the trajectories are straight lines. We can use properties of how the trajectories change, for example after reflecting off a boundary, to reduce the cost of recalculating when the trajectory will next hit a given face of the boundary of a $d$-dimensional polytope from $O(d)$ to $O(1)$.

We provide detailed experiments up to dimension 500, and a comparison
with Hamiltonian Monte Carlo up to dimension 100.  These show that our
method can be an order of magnitude faster, for the same level of
precision, as the current best algorithms using Hamiltonian Monte
Carlo.

\section{Volume estimation algorithms}
\label{sec:vol-estim-algos}
%%i%%%%%%%%%%%%%%%%%%%%%%%%%%%%%%%%%%%%%%%%%%%%%%%%%%%%%%%%%%%%%%%%%%%%%%%%%%%%%%%

Without loss of generality, we consider a polytope $H$ in $d$ dimensions with the origin $0$ strictly inside $H$. We define the polytope $H$ through a $d\times d$ matrix $A$ and a $d$-vector $b$:
\[
    H = \{x ~|~ \forall i,  (Ax)_i \leq b_i\}.
\]
We wish to estimate the volume of $H$,
\[
Vol(H) =\int_H\mbox{d}x.
\]

Our approach is based on the algorithm of \cite{cousins2016practical}.
Consider a sequence of isotropic Gaussians with marginal variances
$\sigma_0^2 < \sigma_1^2 < ... < \sigma_m^2$. Let $f_i$ denote the
density of the Gaussian with variance $\sigma_i$.  We discuss the
choice of the variances below, but \cite{cousins2016practical} assumes
that $\sigma_m$ is sufficiently large so that the Gaussian of variance
$\sigma_m$ is nearly flat on $H$. These assumptions mean that
\begin{eqnarray*}
\int_H f_m(x) dx &=& \left(\frac{1}{\sqrt{2\pi}\sigma_m}\right)^d \int_H \exp\left\{\frac{-1}{2\sigma_m^2} \|x\|^2 \right\} \mbox{d}x \\
&\approx& \left(\frac{1}{\sqrt{2\pi}\sigma_m} \right)^d Vol(H).
\end{eqnarray*}

Thus the  volume of $H$ can be rewritten as:
\begin{eqnarray*}
Vol(H) &\approx& (2\pi\sigma_m^2)^{d/2} \int_H f_m(x) dx \\ 
& = &  (2\pi\sigma_m^2)^{d/2} \int_H f_0(x) dx\   \prod_{i=1}^m \frac{\int_H f_i(x) dx}{\int_H f_{i-1}(x) dx} 
\end{eqnarray*}
We can rewrite each ratio in the product as:
\begin{equation} 
\label{eq:product_term}
    \frac{\int_H f_i(x) dx}{\int_H f_{i-1}(x) dx} = \int_H \frac{f_{i}(x)}{f_{i-1}(x)} \frac{ f_{i-1}(x)}{\int_H f_{i-1}(y) dy} dx,
\end{equation}
which is just the expectation of $\frac{f_i(X)}{f_{i-1}(x)}$ where $X$ is  distributed as a Gaussian with variance $\sigma_{i-1}^2$ restricted to $H$. 
Finally, \cite{cousins2016practical} assumes $\sigma_0$ is sufficiently small that almost all the mass of the Gaussian with variance $\sigma_0^2$ lies within in $H$. This means that $\int_H f_0(x)\mbox{d}x \approx 1$.

This leads to the following approach to estimate $Vol(H)$:
\begin{itemize}
    \item[(1)] Choose $m$, the variances $\sigma^2_0<\cdots<\sigma^0_m$ and Monte Carlo sizes $N_1,\ldots,N_m$.
    \item[(2)] For each $i=1,\ldots,m$, use an MCMC algorithm, or other, to get $N_i$ draws $x^{(i)}_1,\ldots,x^{(i)}_{N_i}$ from a Gaussian with variance $\sigma_{i-1}^2$ restricted to $H$.
    \item[(3)] For each $i=1,\ldots,m$  construct the estimator of (\ref{eq:product_term}) as
    \begin{equation}
    \label{eq:Ri-estim}
    \hat{I}_i= \frac{1}{N_i}\sum_{j=1}^{N_i} \frac{f_i(x^{(i)}_j)}{f_{i-1}(x^{(i)}_j)}.
    \end{equation}
    \item[(4)] Estimate $\log\{ Vol(H)\}$ as
    \begin{equation} \label{eq:estimator}
\frac{d}{2}(2\pi\sigma_m^2) +    \sum_{i=1}^m \log(\hat{I}_i).
    \end{equation}
\end{itemize}

In the original algorithm \cite{CousinVolumeN3}, the number of phases, the variances $\sigma_i$ and the number of steps per phase were chosen deterministically, leading to a $\ostar{d^3}$ complexity. In the practical implementation \cite{cousins2016practical}, a heuristic was added to find a better sequence of Gaussians, reducing the number of steps. In addition, to reduce the number of steps $N_i$, a convergence diagnosis based on a heuristic was added. It uses a sliding window
of  size $W$ (tied to the mixing time of the random walk) to estimate the ratio of Eq. \ref{eq:Ri-estim}. The number of samples $N_i$ is therefore not a fixed number, and tuning $W$ is a non trivial issue~\cite{chevallier2022improved}. Currently the best implementations of this approach \cite{chevallier2022improved}, uses Hamiltonian Monte Carlo \cite{neal2011mcmc} to sample from the restricted Gaussians in step (2). 

In this work, we improve  this algorithm in three respects.

First, in using the first Gaussian, $\sigma_0$ is chosen so that the Gaussian with variance $\sigma_0^2$ has probability mass of between $c_{min}$ and $c_{max}$ within $H$. Practically, we use $c_{min} = 0.1$ and $c_{max} = 0.2$. This introduces an additional term to the estimator -- we need to add an estimate of the log of the probability mass that $f_0$ places with $H$ to (\ref{eq:estimator}). To estimate this probability mass, we sample points from the Gaussian 
and use rejection sampling. %This yields a reliable estimation since the target ratio is between $0.1$ and $0.2$ and is done without using a random walk. 
The advantage of this adaptation is that it allows us to take a much larger variance, $\sigma_0^2$, for the initial Gaussian, and thus a smaller value of $m$.

Second, we remove the stopping criterion on the window size used as a proxy for the number of samples $N_i$. The size of the sliding windows $W$ has to be tuned, which proved complicated to do. Furthermore, the target precision was not reliably obtained. We use a different strategy where a global given computational budget $N$ is used (outside of the tuning phase of the random walks). For each Gaussian of variance $\sigma_i$, we compute $ess_i$ the Effective Sample Size (ESS) per iteration of the random walk using the method described in  \cite[Section 16.4.2]{STAN}. Then we choose the target number of samples $(N_i)_{i\leq m}$ such that each ratio has about the same number of independent samples, i.e. $N_i \times ess_i \approx N_j \times ess_j$ for all $i$ and $j$, and $\sum_{i=1}^m N_i = N$. 

Third, as explained in the next section, we introduce a novel strategy to sample the Gaussians.

\section{Restricted Gaussian sampling using Piecewise Deterministic Markov Processes (PDMP)}
%%i%%%%%%%%%%%%%%%%%%%%%%%%%%%%%%%%%%%%%%%%%%%%%%%%%%%%%%%%%%%%%%%%%%%%%%%%%%%%%%%

\subsection{Bouncy Particle Sampler on Unbounded Space}

We   describe the PDMPs, and proceed with the Bouncy Particle Sampler   on an unbounded space.

\paragraph{PDMP.}
A PDMP, $z_t$, is a continuous time Markov process defined by:
\begin{enumerate}
    \item a deterministic flow $\phi_t(z)$,
    \item a jump rate $\lambda(z)$, and
    \item a jump kernel $q(\cdot|z)$.
\end{enumerate}
The process $z_t$ follows the deterministic flow until a jump event happens. Jumps happen with probability
$\lambda(z(t))dt + o(dt)$ in the interval $[t,t+dt]$. If a jump event happens at time $t$, the process $z_t$ jumps using the jump kernel $q$:
\[
    z_t \sim q(\cdot\mid z_{t^-}),
\]
where $z_{t^-} = \lim_{s\rightarrow t, s < t} z_s$.

\paragraph{Bouncy Particle Sampler.}
For a given target density $\pi$ on $\mathbb{R}^d$, the state space of the Bouncy Particle Sampler is extended by adding a velocity vector, yielding the state space:
\[
    E =  \mathbb{R}^d \times \mathbb{R}^d.
\]
Thus the state of our PDMP is $z=(x,v)$ and the target density becomes $\mu(x,v) = \pi(x) p_v(v)$ where $p_v(\cdot)$ is the density of a multivariate normal of mean $0$ and covariance matrix $I_d$. The Bouncy Particle Sampler is defined as follow:
\begin{enumerate}
    \item $\phi_t(x,v) = (x + tv,v)$,
    \item $\lambda(x,v) = \max(0, - \dotp{\nabla_x (\log \pi)(x)}{v})$, and 
%% Dirac delta
    \item $q(\cdot|z) = \delta_{R(z)}$ with 
    \begin{equation}
        \label{eq:Rxv}
    R(x,v) = \left(x, v - 2 \frac{\dotp{v}{\nabla_x (\log \pi)(x)}}{\|\nabla_x (\log \pi)(x)\|^2}\nabla_x (\log \pi)(x)\right)\end{equation}
\end{enumerate}
Note that the latter formula corresponds to a reflection  of vector $v$
with respect to the gradient of the potential.
In practice, computing the jump times requires simulating a Poisson process of intensity
$\lambda(z_t)$. This is done by sampling $u$ from an exponential law of rate 1, then solving for $t$ the equation
\[
    \int_0^t \lambda(z_s) ds = u.
\]
For a Gaussian target of the form $\pi(x) \propto \exp(-a \|x\|^2)$, finding the event times amounts to solving:
\begin{equation} \label{eq:PoissonRate}
    \int_0^t \max(0,2a (\dotp{x}{v} + s \dotp{v}{v})) ds = u,
\end{equation}
which gives a quadratic equation.

\subsection{Bouncy Particle Sampler Restricted to a Polytope}

The Bouncy Particle Sampler just introduced requires the target
density to be continuous and almost everywhere
differentiable. Therefore restriction of a Gaussian to a polytope
requires adding jumps whenever the process reaches the boundary
\cite{BIERKENS2018148}. We will write $q_b(\cdot\mod z)$ the jump
kernel at the boundary.

%We define the polytope $H$ through a matrix $A$ and a vector $b$:
%\[
%    H = \{x | \forall i,  (Ax)_i \leq b_i\}.
%\]
Our target density is of the form
\[
    \pi(x) \propto \exp(-a \|x\|^2) 1_H(x),
\]
for some constant $a$ that depends on the variance of the Gaussian, and
where $1_H$ is the indicator function of $H$.
At a point $x$ of the boundary $\partial H$, we write $n(x)$ the outward normal,
$\mathcal{V}_x^+ = \{v\in\mathcal{V}| \dotp{n(x)}{v} \geq 0\}$ the set of outgoing velocities,
$\mathcal{V}_x^- = \{v\in\mathcal{V}| \dotp{n(x)}{v} < 0\}$ the set of in-going velocities.
Then the process will target the correct distribution if the jump kernel at the boundary $q_b$ satisfies the following condition~\cite{BIERKENS2018148}:
\begin{equation}
    \int_{\mathcal{V}_x^-} \dotp{n(x)}{u} q_b(x,v|u)du = -\dotp{n(x)}{v},
    \label{eq:boundary}
\end{equation}
where $x$ is on the boundary, $n(x)$ the outward normal at $x$, and $v \in \mathcal{V}^+$ an 
outgoing velocity.
The simplest dynamics that satisfy (\ref{eq:boundary}) are to reflect the trajectory off the boundary, so the new velocity becomes $v'$ with
\begin{equation}
\label{eq:new-velocity}
 v' = v - 2 \frac{\dotp{n}{v}}{\|n\|^2}n,
\end{equation}
with  $n$ the normal to the boundary that was hit.

%Reflecting on the boundary is a valid solution \cite{BIERKENS2018148}, hence  when the trajectory hits the polytope boundary, the velocity is reflected on the boundary. 
Finally, the Bouncy Particle Sampler is not always ergodic \cite{BPSOriginal}. To ensure ergodicity, we add a refresh event with constant rate $\lambda_{refresh}$ \cite{BPSOriginal}. At refresh events, the velocity is resampled from its marginal invariant distribution, the multivariate normal distribution.

\begin{figure}
    %\centering
    \vspace{-1cm}
    \includegraphics[scale = 0.30]{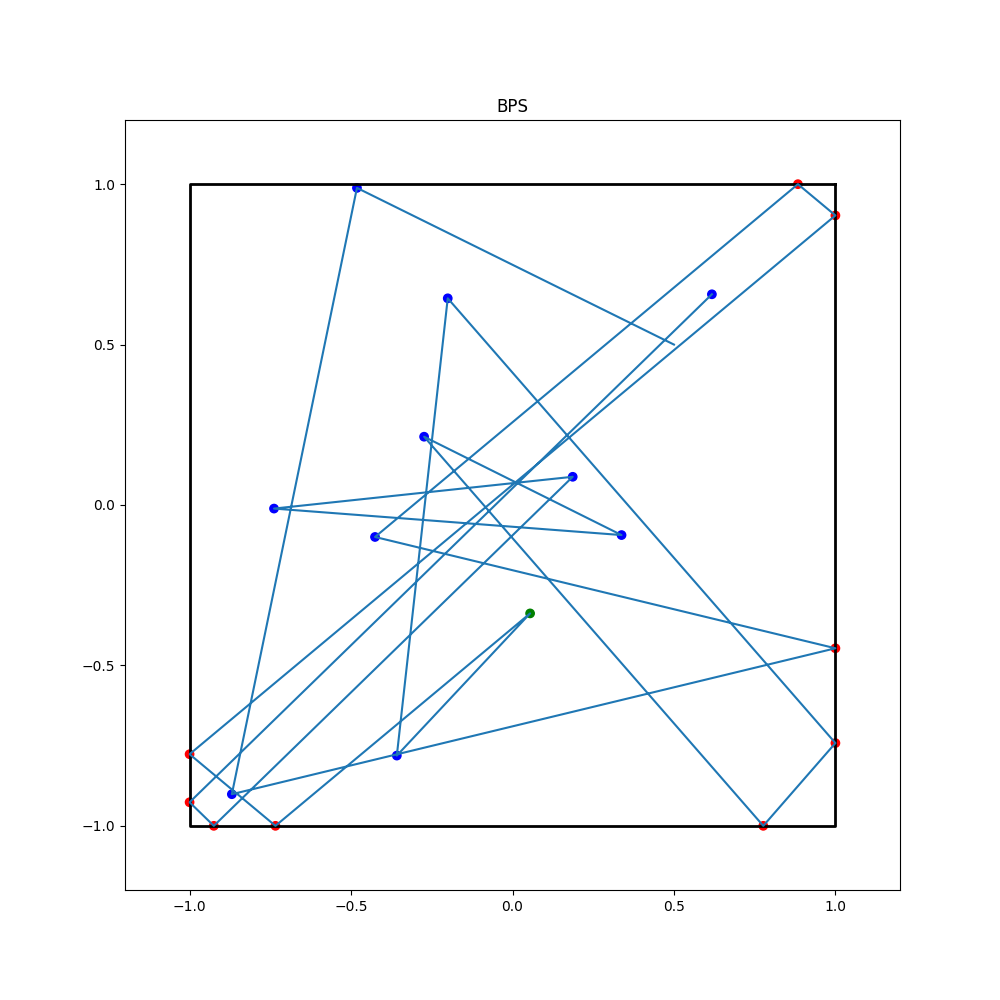}
    \caption{{\bf Example BPS trajectory.} in the 2d cube $[-1,1]^2$, Gaussian of variance $\sigma = 1$. Blue: PDMP jump events, Red: reflections on the boundary, Green: refresh events}
    \label{fig:traj_BPS_example}
\end{figure}

The full algorithm is described in Algorithm \ref{alg:BPS-full}, and an example trajectory can be found in Figure \ref{fig:traj_BPS_example}. 
\begin{algorithm} % enter the algorithm environment
    \begin{algorithmic} % enter the algorithmic environment
        \WHILE{$t \leq t_{max}$}
            \STATE compute $\tau_H$ the intersection time with $H$
            \STATE compute $\tau_{evt}$ the next event time by solving (\ref{eq:PoissonRate}).
            \STATE compute $\tau_{refresh}$ the next refresh event time
            \STATE set $\tau = \min(\tau_H,\tau_{evt},\tau_{refresh})$
            \STATE set $x = x + \tau v$
            \STATE set $t = t + \tau$
            \IF{$\tau = \tau_H$}
                \STATE let $n$ be the normal of the boundary at intersection
                \STATE set $v = v - 2 \frac{\dotp{n}{v}}{\|n\|}n$ (Eq. \ref{eq:new-velocity})
            \ENDIF
            \IF{$\tau = \tau_{evt}$}
                \STATE set $v = R(x,v)$ (Eq. \ref{eq:Rxv})
            \ENDIF
            \IF{$\tau = \tau_{refresh}$}
                \STATE resample $v$
            \ENDIF
        \ENDWHILE
    \end{algorithmic}
    \caption{\bf Bouncy Particle Sampler (BPS).} % give the algorithm a caption
    %\label{alg1} % and a label for \ref{} commands later in the document
    \label{alg:BPS-full}
\end{algorithm}

\begin{comment}
\begin{remark}
It should be noted that
in the case of a flat distribution restricted to a bounded domain, the Bouncy Particle Sampler is the same as
Hamiltonian Monte Carlo with reflections. % which has been extensively described here CITE PAPERS + OUR PAPER. %For Hamiltonian Monte Carlo, it reflections on the boundary was important for the mixing time, and this seems to be the case for BPS too.
\end{remark}
\end{comment}

\subsection{Efficient Implementation}

An important feature of the Bouncy Particle Sampler is that the velocity is constant between events which leads to deterministic trajectories that are straight-lines. This makes it possible to substantially reduce the computational cost of calculating the times at which the trajectory will hit a boundary. By pre-calculating suitable matrix vector products we can reduce the cost of calculating when the process next hits a boundary from $O(dk)$, where $k$ is the number of hyperplanes that defines $H$, to $O(d)$. This efficient implementation is not possible, for Hamiltonian Monte Carlo samplers due to its non-constant velocity.

%Finally, we improve in most cases the computational cost of finding the intersection with $H$ from $d k$ where $d$ is the dimension and $k$ the number of hyperplans to $k$. 
Starting from a point $x$ with velocity $v$, the intersection time $\tau_i$ for hyperplane $i$ is the solution of $(A(x+tv))_i = b_i$, in other words $\tau_i = \frac{b_i - (Ax)_i}{(Av)_i}$. Finding the intersections for all hyperplanes requires computing the products $Ax$ and $Av$, which leads to a complexity of $dk$. However, in our cases, we keep track of the values of $Ax$ and $Av$ and update them without having to fully recompute the matrix vector product.
To update the value of $Ax$ for each event, we use
\[
    A(x + \tau v) = Ax + \tau Av.
\]
Two cases are faced:  

\noindent{\bf Case 1: Reflection  on the boundary.}
First, if the event is a reflection on the boundary with updated velocity $v'$, we have 
\[
    Av' = Av - 2 \frac{\dotp{n}{v}}{\|n\|} An.  
\]
By precomputing $An$ for each hyperplane of the polytope, we can avoid the matrix multiplication.

\noindent{\bf Case 2: PDMP jump event.} The new velocity $v'$ is the reflection of $v$ with respect to the gradient of the potential. In the case of Gaussians, the gradient is colinear with $x$. Hence
$v' = v - 2 \frac{\dotp{x}{v}}{\|x\|}x$, and we can write 
\[
Av' = Av - 2 \frac{\dotp{x}{v}}{\|x\|}Ax.
\]
Since $Ax$ is known, we can also avoid the matrix-vector product.

Finally, note that for a refresh event, we have no choice but to recompute the matrix-vector product. In practice the proportion of events that are of this type is small.

\begin{remark}
In \cite{chalkis2021geometric}, a similar strategy is used for reflection on the boundary, in a setting where there is no PDMP jump event.
\end{remark}

\subsection{Tuning parameters of the Bouncy Particle Sampler}
The Bouncy Particle Sampler has two parameters that requires to be tuned. We describe here the automatic tuning strategy we implemented.

The Bouncy Particle Sampler is currently described as a continuous time process. To extract points from the trajectory $z_t$, we add another Poisson rate $\lambda_{out}$. Each time an event happens with respect to this Poisson process, the current point $z_t$ is passed to the volume computation algorithm. To tune $\lambda_{out}$, we use the following heuristic: to get a sample independent from the starting point, we require 
$d$ events (see Cases 1 and 2 above) to happen. Hence, we automatically tune $\lambda_{out}$ at runtime so that there is on average $d$ events between points passed to the volume algorithm.

Furthermore, as noted before, we have a refresh rate $\lambda_{refresh}$, whose tuning is important for the mixing time. To tune this parameter, we use an optimization procedure described in more details in Section~\ref{sec:exp-tuning}.
\toblack 
%we monitor ESS for all 1-dimensional projections, 

%SCALING LIMIT REF? . If $\lambda_{refresh}$ is too small, BPS will not be very ergodic. If $\lambda_{refresh}$ is too large, BPS will have a diffusion behaviour. To automatically tune the refresh rate, we start with $\lambda_{refresh} = \lambda_{out}$. Then the tuning algorithm tries to find the best value for $\lambda_{refresh}$ starting from there. To do that, we rely on ESS computations. In practice, an ESS is computed on a 1-dimensional projections of the sampled points. We then compute the ESS for the projection on each coordinate, and take the minimum, that we call $ess_{min}$. However, the main problem with BPS is that while it is not ergodic, the 1-dimensional projections will be very good. Hence we compute $ess_{norm}$ the ESS associated to the norm of our samples. Then, if $ess_{min} < ess_{norm}$, we decrease the refresh rate $\lambda_{refresh}$, and if $ess_{min} > ess_{norm}$, we increase the refresh rate.

\section{Implementation and multiprecision}
%%i%%%%%%%%%%%%%%%%%%%%%%%%%%%%%%%%%%%%%%%%%%%%%%%%%%%%%%%%%%%%%%%%%%%%%%%%%%%%%%%

\subsection{Code overview}

Our code is written in C++, using the number types discussed below.
ESS calculations were carried out using the Autoppl library  \url{https://github.com/JamesYang007/autoppl}.

\subsection{Multiprecision}

\noindent{\bf Trajectories escaping the polytope.}
As noted in \cite{chevallier2022improved}, Hamiltonian Monte Carlo trajectories can escape the polytope due to numerical issues -- a  phenomenon becoming  prevalent in high dimension. The solution introduced in \cite{chevallier2022improved} is to recompute a trajectory leaving   the convex by using interval arithmetic with increased precision. 

Since the trajectory escaping the polytope can lead to a completely wrong estimate at best, and a crash of the program at worse, a similar strategy is used in  this work. Assuming the $T_k$ are the output times of the trajectories that are passed to the volume algorithm (i.e. the sequence used by the volume algorithm is $(x_{T_k})_{k\in \mathbb{N}}$ ), the strategy is as follow:
\begin{enumerate}
    \item at time $T_i$, save the state of the Bouncy Particle Sampler: $x_{T_i}$ and $v_{T_i}$, but also the state of the random generator;
    \item compute the trajectory using double precision until time $T_{i+1}$;
    \item if $x_{T_{i+1}}$ is in the polytope, nothing needs to be done;
    \item else: 
    \begin{enumerate}
        \item roll back the state of the Bouncy Particle Sampler to time $T_i$.
        \item increase the precision of real numbers used in the previous step.
        
        \item compute the trajectory until time $T_{i+1}$.
        \item repeat until $x_{T_{i+1}}$ is in the polytope.
    \end{enumerate}
\end{enumerate}
It should also be noticed that  precision cannot be increased indefinitely. 
As a fallback, if the required precision reaches a certain threshold, we go back to time $T_i$ and sample a new velocity.
In the Experiments below, we report on these outcomes using two statistics:
\begin{itemize}
\item $\#M$: the number of times the {\bf else} above is entered.
\item $\#R$: the number of times the precision limit is reached.
\end{itemize}

\noindent{\bf Number types.}
To handle precision refinements, we use the number type  {\tt boost::multiprecision::mpfr\_float}.

\noindent{\bf Numerical values for volumes.}
Multiprecision is also required to compute the   volume, in two guises.
On the one hand, the  exact volume of the polytopes used in our tests cannot be represented using double precision, when increasing the dimension.
On the other hand, high precision floating points must be used when computing the exponential of the sum of the log of the ratios $\sum_{i=1}^m \log(\hat{I}_i)$. 
To handle these difficulties, we use the number type \\
{\tt boost::multiprecision::cpp\_dec\_float},
which makes it possible to specify the number of decimal digits used. 

\section{Experiments}
\label{sec:exp}
%%i%%%%%%%%%%%%%%%%%%%%%%%%%%%%%%%%%%%%%%%%%%%%%%%%%%%%%%%%%%%%%%%%%%%%%%%%%%%%%%%

\subsection{Setup and statistics of interest}
\paragraph{Polytopes.}
We study our algorithm for three polytopes where the exact volume is known:
\begin{itemizep}
    \item The cube, $-1\leq x_i \leq 1$, for $i=1,\ldots,d$.
    \item The standard simplex ($\simplexstd$), $\sum x_i \leq 1$, $x_i \geq 0$.
    \item The isotropic simplex   ($\simplexiso$), a simplex which is also a regular polytope.
\end{itemizep}

\paragraph{Targeting a given  error.}
Assessing the complexity of our algorithm as a function of the
dimension--for a given polytope, requires finding the number of
samples for which a target error on the volume estimate is obtained.  To this end, consider a
minibatch of repeats (24 in our case to exploit brute force
parallelism on our computer), each using $N$ samples.  We run a binary
search on $N$ to obtain the smallest value for which the median error of
the minibatch lies in the interval $4\% \pm 1\%$.  (Nb: to speed up
calculations, the binary search is also exited if the number of
samples between two consecutive runs varies less than 5\%.)

In this experiment, the dimensions $d=50,70,100,140,175,250$ are used
for each polytope.  For a given statistic of interest (the median
running time or the final number of samples $N$), we then perform a
linear regression in {\em log log} scale, to assess the polynomial
scaling of the statistic.

The output of each computation contains
\begin{itemize}
    \item the final estimation of the volume,
    %\item the number of calls to the oracle, \fc{we do not report this for pdmp}
    \item the run-time (in seconds), and
    \item the number of times the precision had to be increased to stay in the polytope.
\end{itemize}
%In item 2, calls to the oracle refer to calculations of the
%intersection of a line and the polytope boundary. The number of calls
%to the oracle can be viewed as a software-independent measure of
%computational cost.

\paragraph{Using a fixed number  of samples.}
To compare PDMP against state-of-the-art volume algorithms based
on Hamiltonian Monte Carlo \cite{chevallier2022improved}, we resort to
experiments at a fixed number of samples. The volume computation is
launched for dimension $d = 100,500$, using sample sizes
 $N=10^5, 10^6, 10^7$ -- the latter for $d=500$ only.

\paragraph{Computer used.} Calculations were run on a desktop DELL Precision 7920 Tower (Intel
Xeon Silver 4214 CPU at 2.20GHz, 64 Gb of RAM), under Linux Fedora
core 34.

\subsection{Tuning}
\label{sec:exp-tuning}
Based on the results in \cite{deligiannidis2018randomized}, heuristically, a small refresh rate will lead to better mixing for individual coordinates of the samples, whereas a higher refresh rate will mean better mixing for functionals such as the norm of the position. In our case, bouncing on the boundary of the polytope can play a similar role to refreshing the velocity. Hence we expect the optimal refresh rate to depend on the variance of the sampled restricted Gaussian.

Thus we compute the ESS for the projection on each coordinate, and take the minimum, which we call $ess_{min}$. Further we compute $ess_{norm}$, the ESS associated to the norm of our samples. If $ess_{min} < ess_{norm}$, we decrease the refresh rate $\lambda_{refresh}$, and if $ess_{min} > ess_{norm}$, we increase the refresh rate.

\begin{figure}[t]
    \centering
    \ifAISTATS
    \includegraphics[width=1\linewidth]{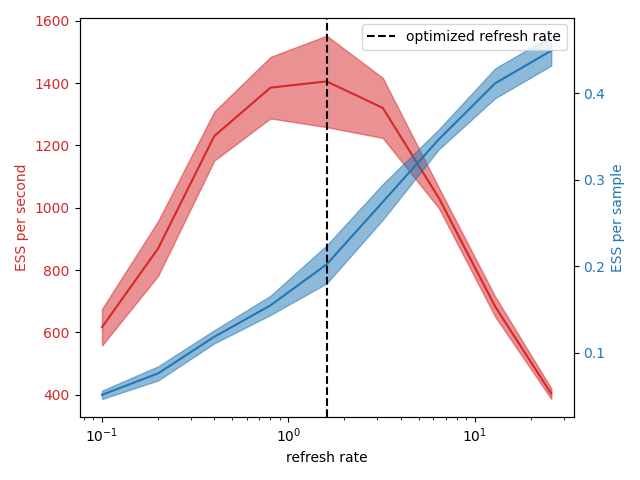}
    \else
    \includegraphics[width=0.8\linewidth]{fig-ess/ESS_optim.png}
    \fi
    \caption{ {\bf ESS optimization of the resfresh rate
      $\lambda_{refresh}$.} Model: cube of dimension $100$. We show in
      red the ESS per second and in blue the ESS per sample around the
      optimized value for $\lambda_{refresh}$. The ESS is computed
      with $10000$ samples. The solid red and blue lines are the avarage of
      the ESS and the envelop represents the standard deviation of the
      ESS.\toblack}
    \label{fig:ess-optim}
\end{figure}

However, refreshing the velocity is expensive, since it requires recomputing the matrix-vector product $A v$, with a cost of $O(d^2)$ instead of $O(d)$ for other events. Thus the new refresh rate is only accepted if it leads to a higher ESS \textit{per second}, with the overall ESS being $min(ess_{min},ess_{norm})$. If not, the tuning is stopped (Fig. \ref{fig:ess-optim}). 

The number of consecutive samples used to evaluate the ESS 
results from a tradeof between the quality of the tuning, which increases
with number of samples, and the time spent tuning the walk with
%%respect to the overall computation time.
In our experiments, the number of consecutive samples is restricted
by runs in dimension 500, which leads us to use 100 consecutive
samples to evaluate the ESS.  %For a higher precision and a higher number of points than what is presented in the paper, a higher number
is advised.

\toblack

\subsection{Results}

\paragraph{PDMP: complexity.}
The experiments at prescribed error rate show a clear polynomial time complexity (Fig. \ref{fig:dim-time-nsamples}), with a scaling around $O(d^{3.5})$ for the three test polytopes (Table \ref{tab:lin-reg}). In particular, we can see that the time complexity is close to the number of samples times $d^2$. This is consistent with intuition, since each sample requires on average $O(d)$ events, and each event (except for refresh events) requires $O(d)$ computations.

\begin{comment}
LINEAR REGRESSION(cube, x=dimension,y=median time -- seconds) on log
 values: slope:3.587761332170939, intercept:-12.741685797073611,
 r_value^2:0.9763517262046408, p_value:0.00021139079820002473,
 std_err:0.27918374102238075

LINEAR REGRESSION(cube, x=dimension,y=num samples) on log values:
 slope:1.751263691655415, intercept:4.598506124581192,
 r_value^2:0.9040572168913266, p_value:0.0035686435963039353,
 std_err:0.28525268794662845

LINEAR REGRESSION(iso simplex, x=dimension,y=median time -- seconds)
 on log values: slope:3.499911867773249,
 intercept:-11.325088614105553, r_value^2:0.9986559501655289,
 p_value:6.777299619430468e-07, std_err:0.06419879132014031

LINEAR REGRESSION(iso simplex, x=dimension,y=num samples) on log
 values: slope:1.7074189763487138, intercept:6.14521853090339,
 r_value^2:0.9962168814517163, p_value:5.373777162567554e-06,
 std_err:0.05260876843295052

LINEAR REGRESSION(std simplex, x=dimension,y=median time -- seconds)
 on log values: slope:3.145590752812805, intercept:-9.607315318763645,
 r_value^2:0.9909996579804986, p_value:3.0468908241922853e-05,
 std_err:0.14988734887602076 No handles with labels found to put in
 legend.

LINEAR REGRESSION(std simplex, x=dimension,y=num samples) on log
 values: slope:1.3418867609531249, intercept:7.949394241680178,
 r_value^2:0.9600903713120952, p_value:0.0006054210902257556,
 std_err:0.13679450713152114
\end{comment}

\begin{table}[!ht] 
\begin{center}
\begin{tabular}{|c |cc | cc|}
\hline
& \multicolumn{2}{c|}{Time} & \multicolumn{2}{c|}{Num. samples}\\
\hline
model & slope & $R^2$ & slope & $R^2$ \\
\hline
%% pdm-ais-prod0
%% cube & 3.58 & 0.97 &  1.75 & 0.90 \\
%% $\simplexiso$ & 3.49 & 0.99 & 1.70 & 0.99 \\
%% $\simplexstd$ & 3.14 & 0.99 & 1.34 & 0.96\\
%% pdm-ais-prod2
cube & 3.77 & 0.96  & 1.94 & 0.88\\
$\simplexiso$ & 3.52 & 1.00 & 1.72 & 0.99\\
$\simplexstd$ & 3.18 & 0.99 & 1.37 & 0.96\\
\hline
\end{tabular}
\end{center}
\caption{{\bf Linear regression in log log scale for the three polytopes.}
First variable regressed: running time (seconds); second variable: number of samples $N$
used to obtain a prescribed error estimate on the volume.
} 
\label{tab:lin-reg} 
\end{table} 
\toblack

\ifAISTATS
\begin{center}[!ht]
\else
\begin{figure}
\fi
\begin{center}
\begin{tabular}{c}
\ifAISTATS
\includegraphics[width=\linewidth]{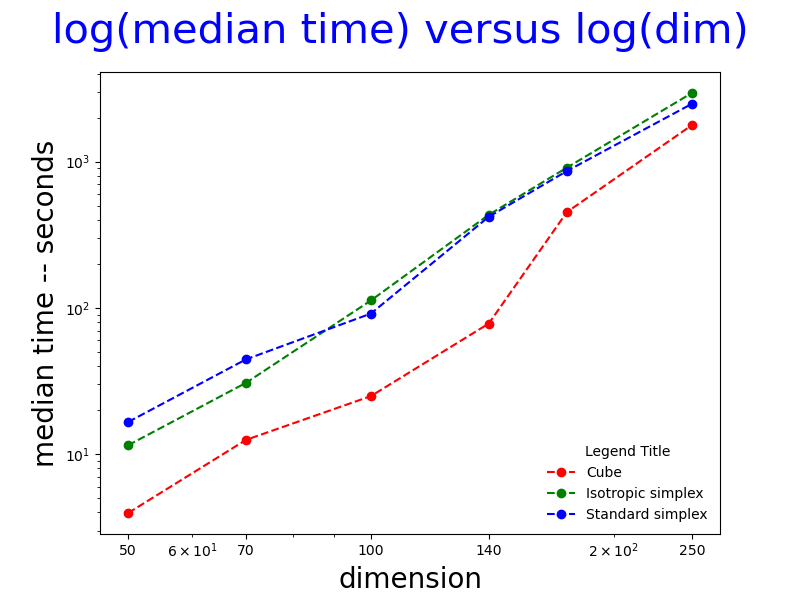}\\
\includegraphics[width=\linewidth]{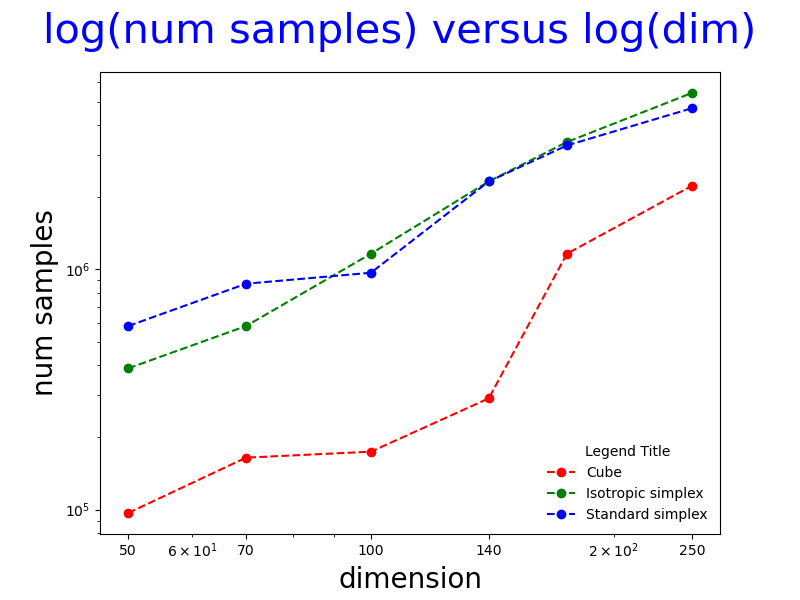}\\
\else
\includegraphics[width=0.8\linewidth]{pdm-ais-prod2/dim-time.png}\\
\includegraphics[width=0.8\linewidth]{pdm-ais-prod2/dim-nsamples.png}\\
\fi
\end{tabular}
\end{center}
\caption{{\bf Complexity of PDMP studied using the smallest number of
    samples achieving a target error estimate for each polytope.}
  Dimensions used: $d=50,70,100,140,175,250$. Plots are in {\em log
    log} scale.  {\bf (Top)} Time as a function of dimension {\bf
    (Bottom)} Number of samples as a function of dimension.  See also
  Table \ref{tab:lin-reg}.  }
\label{fig:dim-time-nsamples} 
\end{figure}

\paragraph{PDMP versus HMC: complexity.}
Hamiltonian Monte Carlo based methods require the specification of a
so-called window size used to determine convergence of the ratio of
(\ref{eq:product_term}) \cite{chevallier2022improved}.  In the sequel,
we use two window sizes:
\begin{itemizep}
\item Low profile comparison: the window size $W=250$ is used to
  compare Hamiltonian Monte Carlo against PDMP with $10^5$ samples --
  short-hand: HMC1;
\item High profile comparison:  the window size 
$W=250+d\sqrt{d}$ is used to compare Hamiltonian Monte Carlo against PDMP with $10^6$ samples -- short-hand: HMC2.
\end{itemizep}
Two facts stand out (Tables \siref{tab:hmc-pdmp} and
\siref{tab:hmc-pdmp2}).  On the one hand, for the three polytopes, the
two algorithms compared yield equivalent error estimates. On the other
hand, PDMP yields a speedup between one and two orders of magnitude
over HMC. To be precise, in dimension $d=100$ e.g.: between $\sim$ 56
and 135 for the low profile comparison, and between $\sim$ 14 and 44
for the high profile comparison. This comparison shows the superiority
of linear trajectories over curved ones in HMC.

As a final analysis, we challenge PDMP with calculations up to
dimension $d=500$, using a fixed number of samples (Tables
\siref{tab:pdm-ais-prod2-d500-vol} and
\siref{tab:pdm-ais-prod2-d500-misc}).  Obtaining a relative error
below 10\% now requires using a number of samples $N$ one order of
magnitude larger. Still, with a running time of the order of hours, such 
large dimensions  remain tractable.
  \toblack

\paragraph{PDMP versus HMC: numerics.}
Another important issue is that of numerical robustness faced by
geometric algorithms in general~\cite{kettner2008classroom}. Along
with the volumes, we monitor the number of multi-precision refinements
-- Table \siref{tab:pdm-ais-prod2-misc}. In sharp contrast with the
method based on Hamiltonian Monte Carlo \cite{chevallier2022improved},
such refinements are not triggered by the PDMP sampler -- the statistic \#M is
null  and therefore so is \#R.  We conjecture that multiprecision is
required with HMC due to intersections between curved trajectories and
the polytope boundary, requiring the (numerically challenging)
computation of an $\arctan$.  

\paragraph{Cooling schedules: using Gaussians versus balls.}
As mentioned in the Introduction, an alternative algorithm uses balls intersected with the polytope instead of Gaussians restricted to the polytope~\cite{chalkis2019practical}. Since this algorithm only requires uniform sampling, the simpler billiard walk sampler can be used \cite{chalkis2019practical}. 
It should be noted than our sampling strategies are essentially the same than the one provided in \cite{chalkis2019practical}, the main difference being that our random walk handle Gaussians while still having a low computational complexity.

Ball cooling essentially amounts to changing the functions
$f_i$ in Eq. (\ref{eq:Ri-estim}) from Gaussian densities to indicator
functions of balls. In that case, the ratio found in
Eq. (\ref{eq:Ri-estim}) is exactly $0$ or $1$. In principle, this should
lead to a higher variance of $\hat{I}_i$ of Eq. (\ref{eq:Ri-estim}) than
for Gaussians, which would be less efficient.  This is consistent
with known theoretical results
\cite{kannan1997random,cousins2015bypassing} which seems to indeed
indicate a greater efficiency of the Gaussian cooling compared to
balls.

Interestingly, the implementation provided in the preprint
\cite{chalkis2019practical} yields better results, which contradicts
theory \cite{kannan1997random,cousins2015bypassing}. This calls for
further analysis in two directions.  The first is the analysis of the
cooling schedule based on a sequence of balls
\cite{chalkis2019practical} versus a sequence of Gaussians
\cite{cousins2016practical}.  The second one is the tuning of the
random walk generating samples (in our case: refresh rate
$\lambda_{refresh}$, output rate $\lambda_{output}$).

We believe that two things might be happening. First, the tuning
strategy used for balls might outperform ours, which is expensive
since it relies on calculating the ESS. Second, we rely on the sequence of Gaussians
provided by \cite{cousins2016practical}, while a new one, possibly
more optimal for balls, is used in \cite{chalkis2019practical}. 

\section{Code availability}
%%The code is available on github:
Visit the github repo \href{https://github.com/augustin-chevallier/PolytopeVolume}{https://github.com/augustin-chevallier/PolytopeVolume}. 

\section{Conclusion}
%%i%%%%%%%%%%%%%%%%%%%%%%%%%%%%%%%%%%%%%%%%%%%%%%%%%%%%%%%%%%%%%%%%%%%%%%%%%%%%%%%

Recent developments for polytope volume calculation algorithms have exploited two strategies, namely using a sequence of Gaussians or a sequence of balls intersecting the polytope. 
Our work  presents improvements for the former, yielding a substantial speedup (more than one order of magnitude on running times) over state-of-the-art  HMC based methods.
On the other hand, recent work in the latter vein has resulted in faster and more accurate results \cite{chalkis2019practical}, an unexpected observation given the theoretical bounds known to date for both classes of methods. 
An interesting avenue for future work will therefore consist of hybridising both approaches. 

For convex bodies with piecewise $C^1$ boundaries, our algorithm could also be applied. While the improvement to the oracle complexity might not apply, the intersection of linear trajectories with the boundary would improve on HMC.  

On the practical aspect, there are a few improvements that could be made in future work. The most obvious one would be to re-use the points sampled at previous phases using importance sampling. The second would be to introduce a way of refreshing velocities that do not require recomputing the matrix-vector product. 

Polytope volume calculations algorithms first underwent  major improvements in the theoretical realm (with bounds on the mixing times but algorithms lacking efficiency in practice), and more recently in the practical realm (with efficient algorithms lacking theoretical bounds). Combining both is clearly an outstanding challenge ahead.

\noindent{\bf Acknowledgments.}  
We thank Sylvain Pion for stimulating discussions on numerical issues.

This work has been partially supported by the French government,
through the 3IA C\^ote d’Azur Investments in the Future project
managed by the National Research Agency (ANR) with the reference
number ANR-19-P3IA-0002, and EPSRC grants EP/R034710/1 and EP/RO18561/1.
\toblack

%Sylvain
%authors autoppl
%Cousins Vemapala matlab code
%Funding bodies.

\bibliographystyle{unsrt}
\bibliography{biblio,local}

\renewcommand{\figurename}{Figure S\hspace{-.225cm}~}
\renewcommand{\tablename}{Table S\hspace{-.225cm}~}
\setcounter{figure}{0}
\setcounter{table}{0}

\ifAISTATS
\onecolumn
\aistatstitle{Efficient computation of the the volume of a polytope in high-dimensions using Piecewise Deterministic Markov Processes: \\
Supplementary Materials}

\section{Results}

\else

\section{Supporting Information}
%%i%%%%%%%%%%%%%%%%%%%%%%%%%%%%%%%%%%%%%%%%%%%%%%%%%%%%%%%%%%%%%%%%%%%%%%%%%%%%%%%

\fi

\subsection{Comparisons at a fixed error rate}
%%ii-%-%-%-%-%-%-%-%-%-%-%-%-%-%-%-%-%-%-%-%-%-%-%-%-%-%-%-%-%-%-%-%-%-%-%-%-%-%-%

\begin{itemize}
\item 
Table \siref{tab:pdm-ais-prod2-vol} 
and Table \siref{tab:pdm-ais-prod2-misc} 
present the full results corresponding 
to the protocol {\em Targeting a given  error} from Section 
\ref{sec:exp}.
\end{itemize}

\begin{table}[H]
\resizebox{\columnwidth}{!}{   %% coati
\begin{tabular}{|r | l l r r| r | r r | r r | r r | r r |}
\hline
N & Algo. & model & $d$ & $\eps$ & $Vol$ & $\min \tilde{V}$ & $\max \tilde{V}$ & $med(\tilde{V})$ & $stdev(\tilde{Vol})$ & $med(\relerrvol)$ & $stdev(\relerrvol)$\\
%$median \tilde{V}$ & ratio & $\max \tilde{V}$ & 
\hline
24 & PDMP-N9.7e+04 & cube & 50 & NA &1.126e+15 & 1.04e+15 & 1.292e+15 & 1.129e+15 & 5.653e+13 &  3.089E-02 &  3.410E-02 \\
\hdashline
24 & PDMP-N1.6e+05 & cube & 70 & NA &1.181e+21 & 1.106e+21 & 1.296e+21 & 1.155e+21 & 5.099e+19 &  3.563E-02 &  2.459E-02 \\
\hdashline
24 & PDMP-N1.7e+05 & cube & 100 & NA &1.268e+30 & 1.125e+30 & 1.423e+30 & 1.254e+30 & 6.966e+28 &  3.075E-02 &  3.619E-02 \\
\hdashline
24 & PDMP-N2.9e+05 & cube & 140 & NA &1.394e+42 & 1.245e+42 & 1.553e+42 & 1.386e+42 & 8.808e+40 &  5.590E-02 &  2.983E-02 \\
\hdashline
24 & PDMP-N1.2e+06 & cube & 175 & NA &4.789e+52 & 4.469e+52 & 5.235e+52 & 4.825e+52 & 2.368e+51 &  4.301E-02 &  2.697E-02 \\
\hdashline
24 & PDMP-N2.2e+06 & cube & 250 & NA &1.809e+75 & 1.662e+75 & 1.968e+75 & 1.81e+75 & 7.275e+73 &  2.300E-02 &  2.593E-02 \\
\hline
24 & PDMP-N3.9e+05 & $\simplexiso$ & 50 & NA &3.421e+21 & 3.173e+21 & 3.65e+21 & 3.419e+21 & 1.127e+20 &  2.462E-02 &  1.968E-02 \\
\hdashline
24 & PDMP-N5.8e+05 & $\simplexiso$ & 70 & NA &1.658e+30 & 1.529e+30 & 1.888e+30 & 1.64e+30 & 7.302e+28 &  3.066E-02 &  2.955E-02 \\
\hdashline
24 & PDMP-N1.2e+06 & $\simplexiso$ & 100 & NA &1.771e+43 & 1.648e+43 & 1.847e+43 & 1.783e+43 & 4.733e+41 &  1.739E-02 &  1.675E-02 \\
\hdashline
24 & PDMP-N2.3e+06 & $\simplexiso$ & 140 & NA &4.167e+60 & 3.855e+60 & 4.441e+60 & 4.223e+60 & 1.73e+59 &  3.069E-02 &  2.139E-02 \\
\hdashline
24 & PDMP-N3.4e+06 & $\simplexiso$ & 175 & NA &6.607e+75 & 5.94e+75 & 6.928e+75 & 6.66e+75 & 2.726e+74 &  2.896E-02 &  2.755E-02 \\
\hdashline
23 & PDMP-N5.4e+06 & $\simplexiso$ & 250 & NA &2.466e+108 & 2.22e+108 & 2.762e+108 & 2.466e+108 & 1.223e+107 &  2.916E-02 &  3.131E-02 \\
\hline
24 & PDMP-N5.8e+05 & $\simplexstd$ & 50 & NA &3.288e-65 & 3.136e-65 & 3.525e-65 & 3.254e-65 & 1.144e-66 &  2.917E-02 &  1.702E-02 \\
\hdashline
24 & PDMP-N8.7e+05 & $\simplexstd$ & 70 & NA &8.348e-101 & 7.864e-101 & 9.148e-101 & 8.437e-101 & 3.563e-102 &  2.716E-02 &  2.651E-02 \\
\hdashline
24 & PDMP-N9.7e+05 & $\simplexstd$ & 100 & NA &1.072e-158 & 9.614e-159 & 1.191e-158 & 1.084e-158 & 5.896e-160 &  2.851E-02 &  3.276E-02 \\
\hdashline
24 & PDMP-N2.3e+06 & $\simplexstd$ & 140 & NA &7.428e-242 & 6.621e-242 & 7.962e-242 & 7.264e-242 & 3.31e-243 &  3.717E-02 &  3.101E-02 \\
\hdashline
24 & PDMP-N3.3e+06 & $\simplexstd$ & 175 & NA &8.893e-319 & 8.029e-319 & 9.465e-319 & 8.672e-319 & 4.164e-320 &  4.609E-02 &  2.545E-02 \\
\hdashline
24 & PDMP-N4.7e+06 & $\simplexstd$ & 250 & NA &3.093e-493 & 2.751e-493 & 3.611e-493 & 3.062e-493 & 2.114e-494 &  5.489E-02 &  3.935E-02 \\

\hline
\end{tabular}}
\caption{
%%{\bf #2}
\captionTableVolume}
\label{tab:pdm-ais-prod2-vol} 
\end{table}

\begin{table}[H]
\resizebox{\columnwidth}{!}{   %% coati
% \begin{tabular}{|r| l l r r | r  r r  r r | r r|}
%% without num of samples
\begin{tabular}{|r| l l  r | r  r r  r r | r r|}
\hline
N & Algo. & model & $d$ & $\eps$ 
%% & med($\#S$) commented out since this number varies a bit
& med($\#O/\#S$) & stdev($\#O/\#S$) 
& med($\#M$) & med($\#R$)
& med(time) & stdev(time)\\
\hline
24 & PDMP-N9.7e+04 & cube & 50 & NA & 5.502E+01 &  1.120E+00 & 0 &0 & 3.980E+00 &  1.349E-01 \\
\hdashline
24 & PDMP-N1.6e+05 & cube & 70 & NA & 7.661E+01 &  1.683E+00 & 0 &0 & 1.257E+01 &  3.556E-01 \\
\hdashline
24 & PDMP-N1.7e+05 & cube & 100 & NA & 1.075E+02 &  2.906E+00 & 0 &0 & 2.497E+01 &  8.915E-01 \\
\hdashline
24 & PDMP-N2.9e+05 & cube & 140 & NA & 1.525E+02 &  4.738E+00 & 0 &0 & 7.820E+01 &  2.079E+00 \\
\hdashline
24 & PDMP-N1.2e+06 & cube & 175 & NA & 1.870E+02 &  6.107E+00 & 0 &0 & 4.523E+02 &  9.362E+00 \\
\hdashline
24 & PDMP-N2.2e+06 & cube & 250 & NA & 2.655E+02 &  6.158E+00 & 0 &0 & 1.774E+03 &  4.370E+01 \\
\hline
24 & PDMP-N3.9e+05 & $\simplexiso$ & 50 & NA & 5.325E+01 &  4.078E-01 & 0 &0 & 1.153E+01 &  1.755E-01 \\
\hdashline
24 & PDMP-N5.8e+05 & $\simplexiso$ & 70 & NA & 7.398E+01 &  7.903E-01 & 0 &0 & 3.078E+01 &  7.514E-01 \\
\hdashline
24 & PDMP-N1.2e+06 & $\simplexiso$ & 100 & NA & 1.049E+02 &  1.349E+00 & 0 &0 & 1.126E+02 &  1.651E+00 \\
\hdashline
24 & PDMP-N2.3e+06 & $\simplexiso$ & 140 & NA & 1.473E+02 &  2.441E+00 & 0 &0 & 4.331E+02 &  5.064E+00 \\
\hdashline
24 & PDMP-N3.4e+06 & $\simplexiso$ & 175 & NA & 1.815E+02 &  2.603E+00 & 0 &0 & 9.097E+02 &  1.406E+01 \\
\hdashline
23 & PDMP-N5.4e+06 & $\simplexiso$ & 250 & NA & 2.594E+02 &  2.725E+00 & 0 &0 & 2.944E+03 &  5.608E+01 \\
\hline
24 & PDMP-N5.8e+05 & $\simplexstd$ & 50 & NA & 5.287E+01 &  4.716E-01 & 0 &0 & 1.663E+01 &  2.769E-01 \\
\hdashline
24 & PDMP-N8.7e+05 & $\simplexstd$ & 70 & NA & 7.310E+01 &  5.935E-01 & 0 &0 & 4.453E+01 &  8.768E-01 \\
\hdashline
24 & PDMP-N9.7e+05 & $\simplexstd$ & 100 & NA & 1.039E+02 &  8.778E-01 & 0 &0 & 9.153E+01 &  1.698E+00 \\
\hdashline
24 & PDMP-N2.3e+06 & $\simplexstd$ & 140 & NA & 1.444E+02 &  1.641E+00 & 0 &0 & 4.197E+02 &  4.341E+00 \\
\hdashline
24 & PDMP-N3.3e+06 & $\simplexstd$ & 175 & NA & 1.799E+02 &  1.238E+00 & 0 &0 & 8.625E+02 &  1.202E+01 \\
\hdashline
24 & PDMP-N4.7e+06 & $\simplexstd$ & 250 & NA & 2.560E+02 &  1.543E+00 & 0 &0 & 2.481E+03 &  2.766E+01 \\

\hline
\end{tabular}
}
\caption{
%%{\bf #2}
\captionTablePDM
} 
\label{tab:pdm-ais-prod2-misc} 
\end{table}

\clearpage
\subsection{Comparisons at a fixed number of samples}
%%ii-%-%-%-%-%-%-%-%-%-%-%-%-%-%-%-%-%-%-%-%-%-%-%-%-%-%-%-%-%-%-%-%-%-%-%-%-%-%-%

\begin{itemize}
\item Tables \siref{tab:hmc-pdmp}  and \siref{tab:hmc-pdmp2}  present a comparison between HMC and PDMP, in dimensions  $d=50$ and $d=100$.

\item Table \siref{tab:pdm-ais-prod2-d500-vol} 
and Table \siref{tab:pdm-ais-prod2-d500-misc} 
present the full results corresponding 
to the protocol {\em Using a fixed number  of samples}  from Section 
\ref{sec:exp}.
\end{itemize}

\begin{table}[!ht] 
\begin{center}
\begin{tabular}{|l r | r | r |}
\hline
Algorithm & Model & $med(\relerrvol)$ & $med(time)$\\
\hline
HMC1 & cube &  3.563E-02 & 1.076E+02 \\
HMC2 & cube &  2.363E-02 & 1.452E+02\\
\hdashline
PDMP-$10^5$ & cube &   2.731E-02   & 4.025E+00\\
PDMP-$10^6$ & cube &   2.248E-02    & 3.583E+01\\
\hline
HMC1 & $\simplexiso$ &   5.191E-02 & 1.164E+02\\
HMC2 & $\simplexiso$ &   3.181E-02 & 1.798E+02\\
\hdashline
PDMP-$10^5$ & $\simplexiso$ &   4.751E-02    & 3.712E+00 \\
PDMP-$10^6$ & $\simplexiso$ &   2.394E-02    & 2.877E+01\\
\hline
HMC1 & $\simplexstd$ &   6.691E-02  & 2.302E+02\\
HMC2 & $\simplexstd$ &   5.715E-02 & 3.655E+02\\
\hdashline
PDMP-$10^5$ & $\simplexstd$ &  4.624E-02 & 3.702E+00\\
PDMP-$10^6$ & $\simplexstd$ &   2.760E-02 & 2.907E+01\\
\hline
\end{tabular}
\end{center}
\caption{{\bf Comparison PDMP versus HMC in dimension $d=50$.}
Results are for cube, isostropic-simplex ($\simplexiso$) and
standard-simplex ($\simplexstd$).  HMC- Hamiltonian Monte Carlo based
methods from \cite{chevallier2022improved}; PDMP -- this work with
sample sizes $N=10^5$ and $N=10^6$.  HMC1: $W = 250$; HMC2: $W = 250+
d^{1.5}$
\label{tab:hmc-pdmp} 
}
\end{table}

\begin{table}[!ht] 
\begin{center}
\begin{tabular}{|l r | r | r |}
\hline
Algorithm & Model & $med(\relerrvol)$ & $med(time)$\\
\hline
HMC1 & cube &  4.221E-02 & 9.122E+02\\
HMC2 & cube &  2.274E-02 & 1.765E+03\\
\hdashline
PDMP-$10^5$ & cube &   4.421E-02 & 1.506E+01\\
PDMP-$10^6$ & cube &   2.104E-02 & 1.261E+02\\
\hline
HMC1 & $\simplexiso$ &   5.782E-02 & 7.285E+02\\
HMC2 & $\simplexiso$ &   3.469E-02 & 1.671E+03\\
\hdashline
PDMP-$10^5$ & $\simplexiso$ &   6.347E-02 & 1.368E+01\\
PDMP-$10^6$ & $\simplexiso$ &   3.324E-02 & 9.637E+01\\
\hline
HMC1 & $\simplexstd$ &   8.130E-02 & 1.755E+03\\
HMC2 & $\simplexstd$ &   4.200E-02 & 4.424E+03\\
\hdashline
PDMP-$10^5$ & $\simplexstd$ &  1.054E-01 & 1.350E+01\\
PDMP-$10^6$ & $\simplexstd$ &   2.856E-02 & 9.465E+01\\
\hline
\end{tabular}
\end{center}
\caption{{\bf Comparison PDMP versus HMC in dimension $d=100$.} Results are for cube, isostropic-simplex ($\simplexiso$) and standard-simplex ($\simplexstd$).
HMC- Hamiltonian Monte Carlo based methods
from \cite{chevallier2022improved}; PDMP -- this work with sample
sizes $N=10^5$ and $N=10^6$.  HMC1: $W = 250$; HMC2: $W = 250+ d^{1.5}$
\label{tab:hmc-pdmp2} 
}
\end{table}

\begin{table}[H]
\resizebox{\columnwidth}{!}{   %% coati
\begin{tabular}{|r | l l r r| r | r r | r r | r r | r r |}
\hline
N & Algo. & model & $d$ & $\eps$ & $Vol$ & $\min \tilde{V}$ & $\max \tilde{V}$ & $med(\tilde{V})$ & $stdev(\tilde{Vol})$ & $med(\relerrvol)$ & $stdev(\relerrvol)$\\
%$median \tilde{V}$ & ratio & $\max \tilde{V}$ & 
\hline
48 & PDMP-N10**5 & cube & 100 & NA &1.268e+30 & 1.08e+30 & 1.505e+30 & 1.253e+30 & 8.121e+28 &  4.421E-02 &  4.029E-02 \\
48 & PDMP-N10**6 & cube & 100 & NA &1.268e+30 & 1.189e+30 & 1.332e+30 & 1.259e+30 & 3.594e+28 &  2.104E-02 &  1.544E-02 \\
\hdashline
48 & PDMP-N10**5 & cube & 500 & NA &3.273e+150 & 1.193e+150 & 6.295e+150 & 2.656e+150 & 1.173e+150 &  2.955E-01 &  2.033E-01 \\
48 & PDMP-N10**6 & cube & 500 & NA &3.273e+150 & 2.289e+150 & 4.084e+150 & 3.197e+150 & 4.064e+149 &  9.948E-02 &  6.933E-02 \\
24 & PDMP-N10**7 & cube & 500 & NA &3.273e+150 & 2.868e+150 & 3.593e+150 & 3.25e+150 & 1.666e+149 &  3.969E-02 &  3.274E-02 \\
\hline
48 & PDMP-N10**5 & $\simplexiso$ & 100 & NA &1.771e+43 & 1.31e+43 & 2.145e+43 & 1.697e+43 & 1.819e+42 &  6.347E-02 &  6.675E-02 \\
48 & PDMP-N10**6 & $\simplexiso$ & 100 & NA &1.771e+43 & 1.586e+43 & 1.948e+43 & 1.754e+43 & 8.567e+41 &  3.324E-02 &  2.850E-02 \\
\hdashline
48 & PDMP-N10**5 & $\simplexiso$ & 500 & NA &9.235e+216 & 1.311e+216 & 2.292e+217 & 5.133e+216 & 4.806e+216 &  5.364E-01 &  3.031E-01 \\
48 & PDMP-N10**6 & $\simplexiso$ & 500 & NA &9.235e+216 & 4.967e+216 & 1.38e+217 & 9.04e+216 & 1.938e+216 &  1.410E-01 &  1.284E-01 \\
24 & PDMP-N10**7 & $\simplexiso$ & 500 & NA &9.235e+216 & 7.16e+216 & 1.123e+217 & 9.113e+216 & 9.188e+215 &  4.407E-02 &  6.582E-02 \\
\hline
48 & PDMP-N10**5 & $\simplexstd$ & 100 & NA &1.072e-158 & 7.873e-159 & 1.537e-158 & 1.081e-158 & 1.752e-159 &  1.054E-01 &  9.916E-02 \\
48 & PDMP-N10**6 & $\simplexstd$ & 100 & NA &1.072e-158 & 9.567e-159 & 1.177e-158 & 1.085e-158 & 5.308e-160 &  2.856E-02 &  2.991E-02 \\
\hdashline
48 & PDMP-N10**5 & $\simplexstd$ & 500 & NA &8.196e-1135 & 1.431e-1135 & 2.249e-1134 & 5.193e-1135 & 4.255e-1135 &  4.676E-01 &  3.004E-01 \\
48 & PDMP-N10**6 & $\simplexstd$ & 500 & NA &8.196e-1135 & 4.838e-1135 & 1.546e-1134 & 7.885e-1135 & 1.859e-1135 &  1.504E-01 &  1.487E-01 \\
24 & PDMP-N10**7 & $\simplexstd$ & 500 & NA &8.196e-1135 & 7.02e-1135 & 1.037e-1134 & 8.099e-1135 & 7.98e-1136 &  7.312E-02 &  6.213E-02 \\

\hline
\end{tabular}}
\caption{
%%{\bf #2}
\captionTableVolume}
\label{tab:pdm-ais-prod2-d500-vol} 
\end{table}

\begin{table}[H]
\resizebox{\columnwidth}{!}{   %% coati
% \begin{tabular}{|r| l l r r | r  r r  r r | r r|}
%% without num of samples
\begin{tabular}{|r| l l  r | r  r r  r r | r r|}
\hline
N & Algo. & model & $d$ & $\eps$ 
%% & med($\#S$) commented out since this number varies a bit
& med($\#O/\#S$) & stdev($\#O/\#S$) 
& med($\#M$) & med($\#R$)
& med(time) & stdev(time)\\
\hline
48 & PDMP-N10**5 & cube & 100 & NA & 1.084E+02 &  3.456E+00 & 0 &0 & 1.506E+01 &  4.787E-01 \\
48 & PDMP-N10**6 & cube & 100 & NA & 1.094E+02 &  3.739E+00 & 0 &0 & 1.261E+02 &  3.464E+00 \\
\hdashline
48 & PDMP-N10**5 & cube & 500 & NA & 5.187E+02 &  5.499E+00 & 0 &0 & 1.127E+03 &  6.796E+01 \\
48 & PDMP-N10**6 & cube & 500 & NA & 5.155E+02 &  4.751E+00 & 0 &0 & 6.595E+03 &  6.420E+02 \\
24 & PDMP-N10**7 & cube & 500 & NA & 5.123E+02 &  5.283E+00 & 0 &0 & 5.729E+04 &  1.107E+04 \\
\hline
48 & PDMP-N10**5 & $\simplexiso$ & 100 & NA & 1.070E+02 &  1.309E+00 & 0 &0 & 1.368E+01 &  3.791E-01 \\
48 & PDMP-N10**6 & $\simplexiso$ & 100 & NA & 1.047E+02 &  1.054E+00 & 0 &0 & 9.637E+01 &  1.424E+00 \\
\hdashline
48 & PDMP-N10**5 & $\simplexiso$ & 500 & NA & 5.240E+02 &  3.540E+00 & 0 &0 & 7.503E+02 &  4.174E+01 \\
48 & PDMP-N10**6 & $\simplexiso$ & 500 & NA & 5.114E+02 &  2.122E+00 & 0 &0 & 3.112E+03 &  2.659E+02 \\
24 & PDMP-N10**7 & $\simplexiso$ & 500 & NA & 5.092E+02 &  2.132E+00 & 0 &0 & 2.328E+04 &  2.978E+03 \\
\hline
48 & PDMP-N10**5 & $\simplexstd$ & 100 & NA & 1.061E+02 &  1.022E+00 & 0 &0 & 1.350E+01 &  2.935E-01 \\
48 & PDMP-N10**6 & $\simplexstd$ & 100 & NA & 1.037E+02 &  6.735E-01 & 0 &0 & 9.465E+01 &  9.332E-01 \\
\hdashline
48 & PDMP-N10**5 & $\simplexstd$ & 500 & NA & 5.179E+02 &  3.678E+00 & 0 &0 & 7.727E+02 &  6.141E+01 \\
48 & PDMP-N10**6 & $\simplexstd$ & 500 & NA & 5.053E+02 &  2.175E+00 & 0 &0 & 3.175E+03 &  1.058E+02 \\
24 & PDMP-N10**7 & $\simplexstd$ & 500 & NA & 5.050E+02 &  1.624E+00 & 0 &0 & 2.744E+04 &  3.650E+03 \\

\hline
\end{tabular}
}
\caption{
%%{\bf #2}
\captionTablePDM
} 
\label{tab:pdm-ais-prod2-d500-misc} 
\end{table}

\toblack

\end{document}